\documentclass[aps,pra,twocolumn,superscriptaddress,amsmath,amssymb]{revtex4-2}

\usepackage{bm}
\usepackage{graphicx} 
\usepackage[mathlines]{lineno}
\usepackage[colorlinks=true,linkcolor=blue]{hyperref}
\usepackage{amssymb,epsfig}
\usepackage{color}

\begin{document}
\title{Vibration induced transparency and absorption with two ion ensembles in a linear trap}
\author{Wenjun Shao}
\email{shaowenjun@westlake.edu.cn}
\affiliation{School of Integrated Circuits, and School of IoT Engineering, Wuxi Taihu University, Wuxi 214064, China}
\affiliation{Taihu Institute of Key Technologies for Integrated Circuits, and Provincial Key (Construction) Laboratory of Intelligent Internet of Things Technology and Applications in Universities, Wuxi Taihu University, Wuxi 214064,  China}
\affiliation{ Department of Physics, School of Science, Westlake University, Hangzhou 310030, China}
\affiliation{ Institute of Natural Sciences, Westlake Institute for Advanced Study, Hangzhou 310024, China}

\author{Jian Li}
\email{lijian@westlake.edu.cn}
\affiliation{ Department of Physics, School of Science, Westlake University, Hangzhou 310030, China}
\affiliation{ Institute of Natural Sciences, Westlake Institute for Advanced Study, Hangzhou 310024, China}

\begin{abstract}
We study the spectra of collective low excitations of two atomic ion ensembles which are confined in a liner trap by addressing lases. 
When the left ensemble is driven by an external optical field, its corresponding response spectrum to the incident optical light shows a vibration-induced-transparency phenomenon when the detuning of the laser addressing the ion is tuned to the first red sideband.
In the case of the detuning tuned to the first blue sideband, the response spectrum shows a conversion from  the absorption peak to the transparency window. 
Furthermore, we investigate the fluctuation spectra of the collective excitation modes of ion ensemble and show the similar phenomena.  
\end{abstract}

\maketitle
\section{Introduction}

In the realm of optics and photonics, the dual phenomena of transparency and absorption have long been themes of intense research. Over the years, various mechanisms have been discovered that can manipulate these processes, leading to significant advancements in the understanding of light-matter interactions \cite{Weiner, Kockum}.  
The transparency and absorption phenomena have not only deepened our understanding of the fundamental principles of quantum optics \cite{Scully} but also paved the way for innovative applications in optical communication, sensing, and information processing.

For example, electromagnetically induced transparency (EIT) and absorption (EIA) have garnered significant attention due to their ability to control the optical properties of materials \cite{Akulshin}. EIT \cite{Lukin, EIT, EIT2}, in particular, enables a normally opaque medium to become transparent under specific conditions, while its counterpart, EIA \cite{EIA}, leads to enhanced absorption. These phenomena are attributed to the destructive and constructive interference of quantum pathways, respectively. 
Nagnon-induced transparency (MIT) \cite{BWang, Ullah} or absorption (MIA) \cite{FWang}, a relatively new field, explores the interaction between light and magnons (spin excitations), which offers a means for controlling light with magnetic materials. 
This interplay has significant implications for the development of spintronic and magnonic devices, where the manipulation of magnons can be used to process and transmit information. 
Complementary to these mechanisms, optomechanically induced transparency (OMIT) and absorption (OMIA) \cite{Hou, Zhang, T. Wang} further expand the scope of light-matter interactions by introducing a mechanical degree of freedom. 
In such systems, the interaction between light and a mechanical oscillator \cite{Aspelmeyer}, such as a membrane or a nanorod, gives rise to transparency or absorption. 
This coupling between optical and mechanical modes has potential applications in quantum information processing \cite{YU} and precision measurements \cite{Childs}. 
Vibration induced transparency (VIT), a related but less explored phenomenon, explores the influence of vibrational modes, such as phonons in solids and collective vibration of trapped ions, on the optical characteristics of materials \cite{Radeonychev}.
Shao \textit{et al} investigate the ionic VIT 
in a Paul trap with a single two-level ion driven by two orthogonal laser beams  \cite{shaolpl}, while the ionic vibration induced absorption (VIA) phenomenon is absent. 
Here we explore the VIT and VIA phenomenon in a linear trap by replacing a single two-level ion with two ion ensembles.

In this paper, we investigate a model system analogous to one in which the collective low excitations of two atomic ensembles are coupled to a single-mode cavity field \cite{Turek}.
In our model, two ion ensembles are confined by addressing lases in a liner trap, with one of both driven by an additional laser.
The trapped ions here are assumed to have collective spatial vibrations and restricted to the joint motional ground state \cite{Cirac}.
By bosonizing the low excitations of two trapped ion ensembles and choosing the detuning of the addressing laser from the sideband, we will get different couplings between the quantized vibrational mode and two bosonic modes.
From the quantum Langevin equations, the steady-state response intensities of two ensembles show different phenomena in different conditions, e.g.,  the harmonic vibration induced transparency would appear in the red-detuning case; 
however, for the driven ensemble in the blue-detuning case, the VIT would converts into VIA under proper conditions.
To confirm the above predictions based on a simple model, we also calculate the fluctuation spectra, which display similar phenomena.

\section{Model}

The model under consideration is a string of ions confined in a linear rf trap, where the addressing lasers are tuned sufficiently close to the sidebands, thus we can neglect all other vibrational modes and concentrate on one collective degree of vibrational excitation of the ions \cite{MS1, MS2}. 
Here the trapped two-level ions are divide into two ensembles (labelled with $ y=a$ or $b $), the left ion ensemble ($a $) is driven by a classical external field with frequency $ \omega_f $, amplitude $  \varepsilon $. Then the Hamiltonian of the present system reads ($ \hbar=1 $)
\begin{equation*}
H =  H_1+H_2,
\end{equation*}
\begin{equation}
H_1= \nu \left(  c^{\dagger}c +1/2\right)  + \omega_{eg} \sum_{i} \sigma^i_z/2, \label{H}
\end{equation}
\begin{equation*}
H_2=  \varepsilon \sum^{N_a}_{i=1} \sigma^i_{+} e^{-i\omega_ft} + \sum^{N_a+N_b}_{i=1} \Omega_i \sigma^i_+ e^{i\eta_i \left( c + c^{\dagger} \right) -i\omega_i t} +  H.c.,
\end{equation*}
where $ c^{\dagger} $ and $  c $ are the ladder operators of the vibrational mode and $ \nu $ is the frequency, and $ \hbar \omega_{eg} $ is the energy difference between the ionic internal states $ e $ and $ g $. 
Pauli matrices $ \sigma_i $ represent the internal degrees of freedom for the $ i $th ion in the left ensemble ($ a $)  or the right ensemble  ($ b $) with the number of ions  $ N_a/N_b $,
and the frequency $ \omega_i $ and Rabi frequency $ \Omega_i $ of the laser addressing any $ i $th ion are assumed to be the same, together with the Lamb-Dicke parameter $ \eta_i $.

Considering an ion trap operating in the Lamb-Dicke limit, i.e., $ \eta_i \ll 1 $, we use an expansion of $ H_2 $ to the first order in $ \eta_i $ during the analytical calculations \cite{Leibfried}.  
To simplify the model Hamiltonian, we redefine the new operators of collective excitation modes for the two ion ensembles ($ y=a, b $) \cite{Turek, Liu, Jin}:
\begin{equation}
 Y^{\dagger} = \dfrac{1}{\sqrt{N_y}} \sum^{N_y}_{i=1} \sigma^i_{+} \qquad (Y=A, B).	
\end{equation}
In the low-excitation limit with large $ N_y $, the new collective operators satisfy the  bosonic commutation relations
$ \left[Y,Y^{\dagger} \right] \approx  1$, 
$   \left[A,B\right] =	\left[A,B^{\dagger} \right]=0 $,
and further 
$ \sum^{N_y}_{i=1} \sigma^i_{z} =2 Y^{\dagger}Y-N_y$. Then, Hamiltonian \eqref{H} can be rewritten 
 as 
\begin{equation}
\begin{split}
H  = &  \nu  c^{\dagger}c + \omega_{eg}  \sum_{Y=A,B}  Y^{\dagger}Y  +  \chi  A^{\dagger}  e^{-i\omega_ft}  \\
& +  \sum_{Y=A,B}  Y^{\dagger} \left[ G_{Y} + ig_Y \left( c + c^{\dagger} \right) \right] e^{-i\omega_i t}   + H.c. 
\end{split} \label{H_new}
\end{equation}
with $ \chi = \sqrt{N_a} \varepsilon$, $ G_Y=\sqrt{N_y}\Omega_i $  and $ g_Y= \eta_i G_Y $. 
During the derivation the constant terms $ \nu/2 $ and $ - \omega_{eg} N_y/2$ have been neglected since they do not affect the results in the context.

In the following we will study two different cases in which the frequency of the addressing laser, $ \omega_i $, is respectively tuned to the first red sideband or the first blue sideband of the ionic internal transition frequency $ \omega_{eg} $ \cite{Leibfried}.  
In the red-detuning case, when the frequency of the addressing laser beam is tuned to the first red sideband of the ionic transition, i.e., the frequency of the addressing laser satisfies $ \omega_i=\omega_{eg} -\nu $.  
In the interaction picture with respect to $ H_0= \omega_{f} \left(  A^{\dagger}A +  B^{\dagger}B \right) +\left(\omega_f-\omega_i \right) c^{\dagger}c  $, the interaction Hamiltonian after the rotating-wave approximation (RWA) is given in time-independent form as
\begin{equation}
\begin{split}\label{Hr}
H_r = & \Delta  \left(  A^{\dagger}A +  B^{\dagger}B + c^{\dagger}c\right) + \chi \left(  A^{\dagger}+   A \right) \\
& + ig_A   \left(  A^{\dagger}  c -  A c^{\dagger}  \right) + ig_B \left(   B^{\dagger}c -  B c^{\dagger} \right) ,
\end{split}
\end{equation}
 where the detuning $ \Delta= \omega_{eg}-\omega_f $. 
In the blue-detuning case, the frequency of the addressing laser is set to  $ \omega_i=\omega_{eg} + \nu $.
Making a unitary transformation with $ H_0= \omega_{f} \left(  A^{\dagger}A +  B^{\dagger}B \right) +\left(\omega_i- \omega_f\right) c^{\dagger}c  $, 
 one can obtain the interaction Hamiltonian  
\begin{equation}
\begin{split}\label{Hb}
H_b = & \Delta  \left(  A^{\dagger}A +  B^{\dagger}B - c^{\dagger}c\right) + \chi \left(  A^{\dagger}+   A \right) \\
& + ig_A   \left(  A^{\dagger}  c^{\dagger} -  A c  \right) + ig_B \left(   B^{\dagger}c^{\dagger} -  B c \right)  
\end{split}
\end{equation}
after RWA.

\section{RESPONSE SPECTRUM}

In the considered model the external driving field can be considered as a probe field which is incident from the addressing lasers. Thus these two fields do not disturb each other. Next we study the response spectra of the ionic collective excitation modes to the driving in the red-detuning and blue-detuning cases, respectively.

For the red-detuning case, the quantum Langevin equations based on Eq. \eqref{Hr}  can be written as
\begin{align}
\dot{c} =& -i \Delta c-g_AA-g_BB-\kappa c+\sqrt{2\kappa}c_{in}(t), \label{dotc}\\
\dot{A} =& -i \Delta A-i \chi + g_Ac-\gamma_A A+ \sqrt{2\gamma_A}A_{in}(t),\\
\dot{B} =& -i \Delta B+g_Bc-\gamma_B B + \sqrt{2\gamma_B}B_{in}(t), \label{dotB}
\end{align}
where $ \kappa $ is the heating rate for vibrational motion and $ \gamma_A $ and  $ \gamma_B $ are the decay rate of collective modes A and B instead of the single-ion decay rate, which scale linearly with the numbers of the ions. 
Here the operators $ X_{in}  $ ($ X=A,B,c $) denote the corresponding noises with the vanishing average values, i.e., $ \langle X_{in}\rangle =  0 $. The initial system-bath state is uncorrelated under the first Markov approximation according to fluctuation-dissipation theorem \cite{Gardiner}.  
And the state of each environment is weakly affected by the system and is described by thermal states.
These noise operators satisfy the following fluctuation relations: 
$ \langle X_{in} (t) X^{\dagger}_{in} (t') \rangle = [N(\omega_x)+1]\delta (t-t')$, $ \langle X^{\dagger}_{in} (t) X_{in} (t') \rangle =  N(\omega_x) \delta (t-t')$, where $ N(\omega_x)= [\text{exp}(  \omega_x/k_BT) -1 ]^{-1}$  ($ \omega_x= \omega_{eg}, \nu$)
are, respectively, the equilibrium mean numbers of the ionic collective modes and the vibrational mode at temperature T, with $ k_B $ being Boltzmann constant \cite{FWang}.
The steady-state values of the system are given by
\begin{align}
A_s	& =\langle A \rangle =	-\dfrac{\chi F_A}{\Delta-i\gamma_A},	\label{A_s}\\
B_s	& =\langle B \rangle = -\dfrac{\chi f_af_b}{\Delta_{\text{eff}}-i\kappa_{\text{eff}}}, \label{B_s}\\
c_s & = \langle c \rangle =	-\dfrac{i\chi f_a}{ \Delta_{\text{eff}}-i\kappa_{\text{eff}}}, \label{c_s}
\end{align}
where 
\begin{equation*}
F_A=1+\dfrac{g_Af_a}{\Delta_{\text{eff}}-i\kappa_{\text{eff}}}
\end{equation*}   
is the modified factor of the coupling coefficient between the left ion ensemble and external driving field,
\begin{equation*}
\kappa_{\text{eff}} =\kappa + \dfrac{g_A^2 \gamma_A}{\Delta^2+\gamma_A^2} + \dfrac{g_B^2 \gamma_B}{\Delta^2+\gamma_B^2} 
\end{equation*}
and
\begin{equation*}
\Delta_{\text{eff}}=\Delta\left( 1-\dfrac{g_A^2 }{\Delta^2+\gamma_A^2} - \dfrac{g_B^2 }{\Delta^2+\gamma_B^2} \right) 
\end{equation*} 
are respectively the effective heating rate and detuning, with $  f_a=\dfrac{ g_A}{\Delta-i\gamma_A } $ and $  f_b=\dfrac{ g_B}{\Delta-i\gamma_B }  $.

\begin{figure}[bp]
\centering
\includegraphics[width=0.495\textwidth]{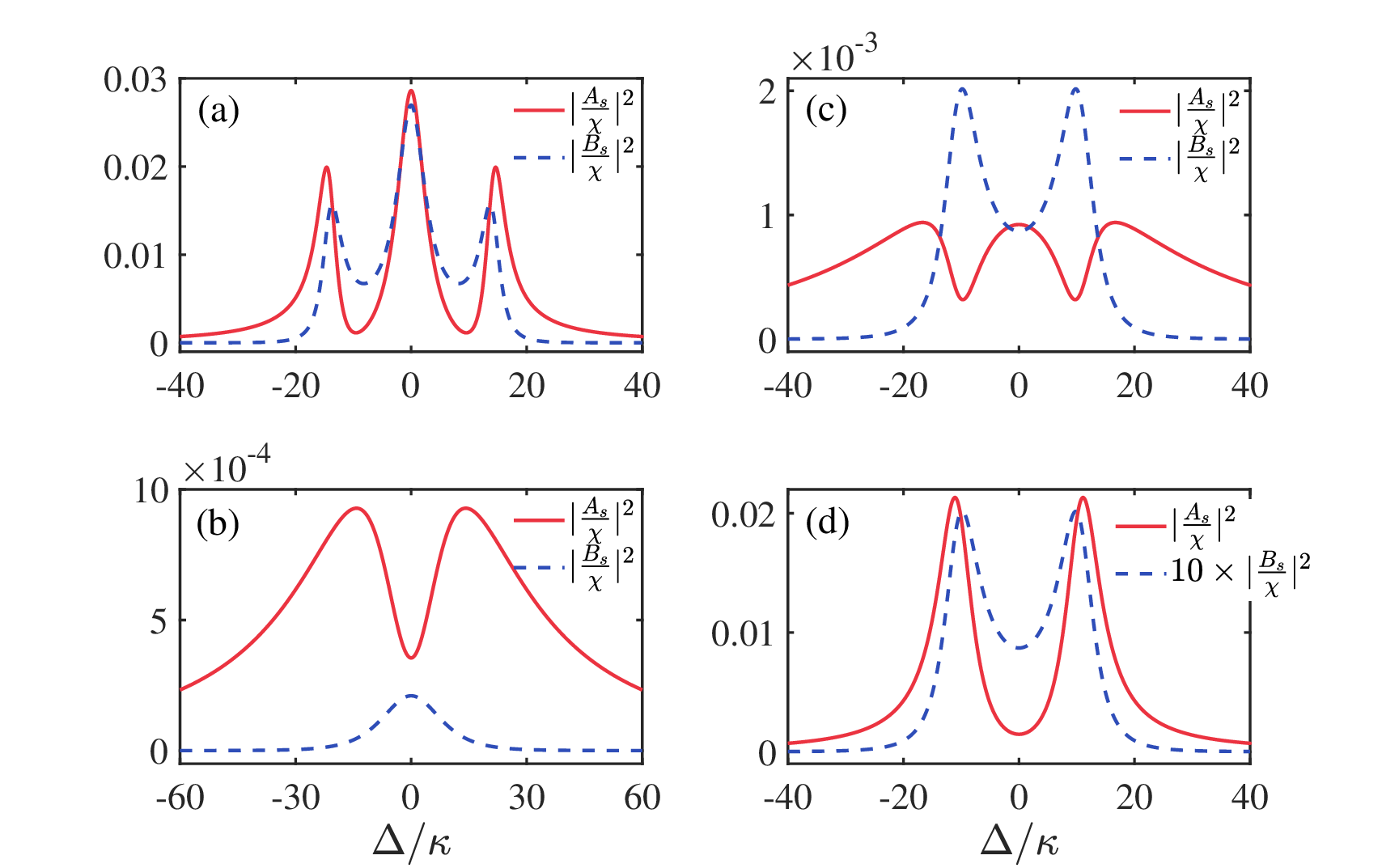} 
\caption{The response intensities of ion ensembles $\vert A_s/\chi \vert^2  $ (red solid curve) and $\vert B_s/\chi \vert^2  $ (blue dashed curve) in arbitrary units as a function of the detuning $ \Delta $ according to Eqs. \eqref{A_s} and \eqref{B_s} with $ g_A=g_B=10 $ but with different decay rates: (a) $\gamma_A =3  $, $\gamma_B =3 $; (b) $\gamma_A =30 $, $\gamma_B =30 $; (c) $\gamma_A =30  $, $\gamma_B =3 $; (d) $\gamma_A =3  $, $\gamma_B =30 $. All the frequencies are in units of $ \kappa $.}
\label{VIT2}
\end{figure}

It can be found that the steady-state values of all the three bosonic modes are proportional to  $ \chi $ from Eqs. \eqref{A_s}--\eqref{c_s}. 
In the following, we will investigate the steady-state response spectra (mean excitation populations $ \vert A_s \vert^2 $ and $ \vert B_s \vert^2 $) of the two collective excitation modes of ion ensembles. 
In this case, the results show some similarities to those of Ref. \cite{Turek}, which studies the spectra of collective low excitations of two trapped ensembles coupled with a single-mode cavity field.

When the number of ions $ N_a/N_b=1 $, e.g., $ g_A = g_B = 10 \kappa$ in Fig. \ref{VIT2},  we plot the response spectra of the ensembles as a function of the scaled detuning $ \Delta/\kappa $.  
If both the decay rates of the collective excitation modes are not so large, e.g, $\gamma_A = \gamma_B =3 \kappa$ ($<g_A$,  $ g_B $), the response spectra of both ensembles appear with two pronounced VIT windows as shown in Fig.  \ref{VIT2}(a). 
However, if the decay rates of the ensembles are very large, e.g,  $\gamma_A = \gamma_B =30 \kappa$ ($>g_A$,  $ g_B $) in Fig. \ref{VIT2}(b), the response of the driven ensemble happens with only one VIT window and the one of the right ensemble appears without any VIT window. 
In the condition of $\gamma_A  =30 \kappa$ and $\gamma_B =3 \kappa$ in Fig. \ref{VIT2}(c), the driven ensemble responds with a two-window VIT phenomenon and the right one with a one-window VIT phenomenon. 
As shown  in Fig. \ref{VIT2}(d) with $\gamma_A  =3 \kappa$ and $\gamma_B =30 \kappa$, the response spectra of both ensembles show one VIT window with the right one a weak VIT response.

\begin{figure}[bp]
\centering
\includegraphics[width=0.51\textwidth]{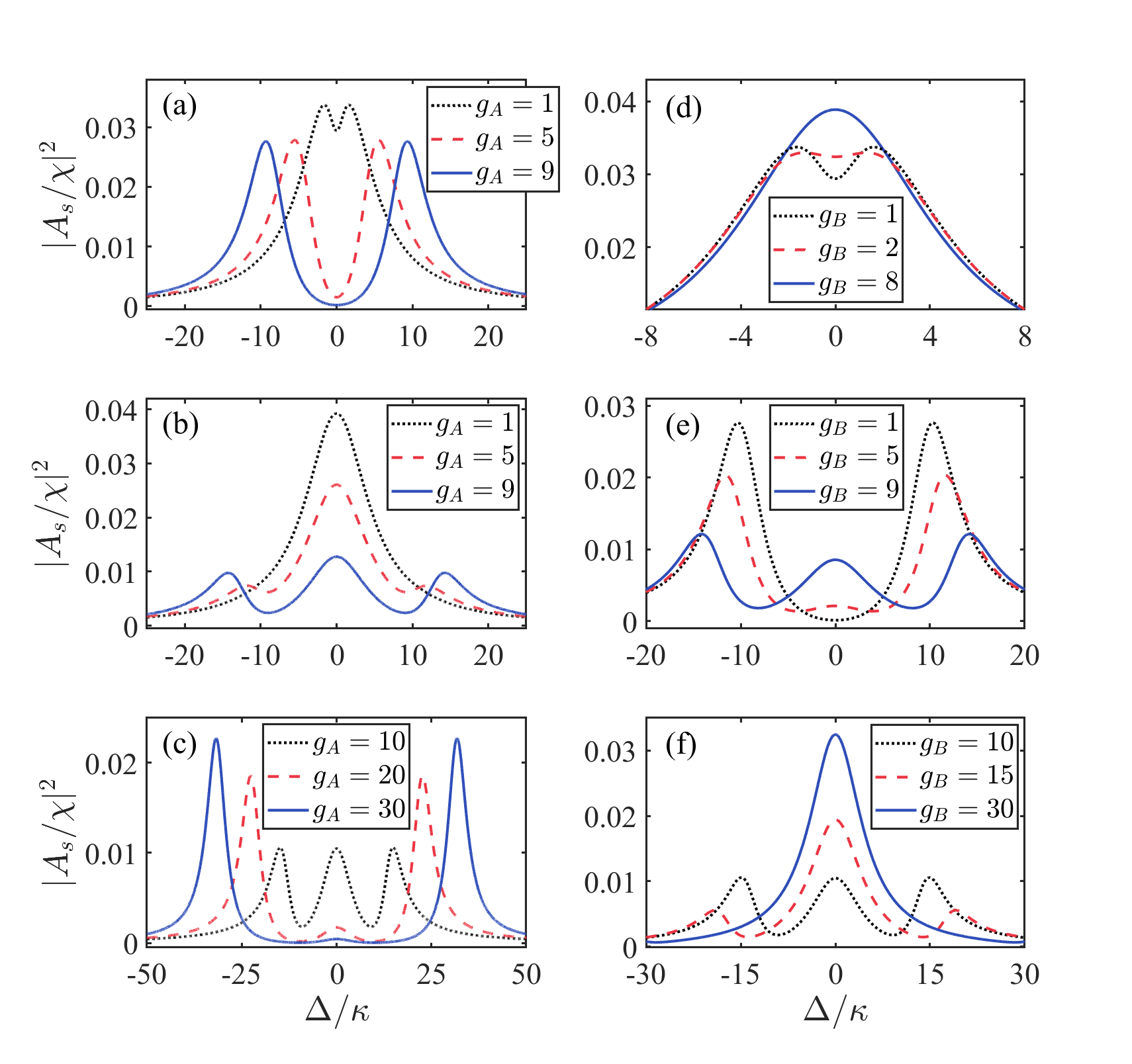} 
\caption{The response intensities of ion ensemble $\vert A_s/\chi \vert^2  $  as a function of the detuning  $ \Delta $ for (a) $ g_B=1 $, (b,c) $ g_B= 10 $ with different coupling strengths $ g_A $, or for (d) $ g_A= 1 $, (e,f) $ g_A= 10 $ with different coupling strengths $ g_B $. 
Here  $\gamma_A = \gamma_B = 5 $ (in units of $ \kappa $).  }
\label{VIT3}
\end{figure}
 
Furthermore, to investigate the effect of the different coupling strengths $ g_A $ and $ g_B $,  in Fig. \ref{VIT3} we plot the response spectra of left ensemble as a function of  $\Delta/\kappa $ for the cases of fixed decay rates  (e.g., $\gamma_A = \gamma_B =5 \kappa $).
From Fig. \ref{VIT3}(a) with fixed $ g_B/\kappa= 1 $,  we find the response intensity $ \vert A_s /\chi \vert^2 $ appears with VIT phenomenon where the dip of the transparency window becomes deeper and the corresponding width becomes wider with the increase of $ g_A $.  
While as shown in Fig. \ref{VIT3}(b,c) with  fixed $ g_B/\kappa= 10 $, the  Lorentz peak in the case of weak-coupling strength $ g_A $ (e.g., $ g_A \sim 1 \kappa$) splits into two VIT windows  in the case of strong-coupling strength $ g_A $ (e.g., $ g_A\sim 10 \kappa$), and then evolves into one VIT window in the case of ultra- and deep-strong-coupling strength $ g_A $ (e.g., $ g_A \geq 30 \kappa$).
For the condition of fixed $ g_A/\kappa= 1 $, it can be seen from Fig. \ref{VIT3}(d) that a mall dip in the  Lorentz peak disappears with the increase of $ g_B $.  
In Fig. \ref{VIT3}(e,f) with fixed $ g_A/\kappa= 10 $,  it  shows that the single VIT window in the case of weak-coupling strength $ g_B $ (e.g., $ g_B\sim 1 \kappa$) splits into the double window in the case of strong-coupling strength $ g_B $ (e.g., $ g_B\sim 10 \kappa$), and eventually becomes a Lorentz peak in the case of ultra- and deep-strong-coupling strength $ g_B $ (e.g., $ g_B \geq 30 \kappa$).
Thus under certain conditions, the VIT windows can be controlled by the amount coupling coefficients.

Similarly, in the blue-detuning case the quantum Langevin equations based on Eq. \eqref{Hb} are given by
\begin{align}
\dot{c} =& i \Delta c + g_AA^{\dagger}+ g_BB^{\dagger}-\kappa c+ \sqrt{2\kappa}c_{in}(t),\\
\dot{A} =& -i \Delta A-i \chi + g_Ac^{\dagger} - \gamma_A A+ \sqrt{2\gamma_A}A_{in}(t),\\
\dot{B} =& -i \Delta B+ g_Bc^{\dagger} -\gamma_B B + \sqrt{2\gamma_B}B_{in}(t).
\end{align}
By using the Hermitian property of the operators, the steady-state solutions are   
\begin{align}
A'_s &  = -\dfrac{\chi F'_A}{\Delta-i\gamma_A},	\\
B'_s &  =  \dfrac{\chi f_af_b}{\Delta'_{\text{eff}}-i\kappa'_{\text{eff}}},\\
c'_s &  =	-\dfrac{i\chi f^*_a}{ \Delta'_{\text{eff}} + i\kappa'_{\text{eff}}},
\end{align}
where
\begin{align*}
F'_A  &= 1-\dfrac{g_Af_a}{\Delta'_{\text{eff}}-i\kappa'_{\text{eff}}}, \\
\kappa'_{\text{eff}}   &= 2\kappa -  \kappa_{\text{eff}} =   \kappa-  \dfrac{g_A^2 \gamma_A}{\Delta^2+\gamma_A^2} - \dfrac{g_B^2 \gamma_B}{\Delta^2+\gamma_B^2},\\
\Delta'_{\text{eff}} &  =2\Delta-\Delta_{\text{eff}} = \Delta \left( 1+\dfrac{g_A^2 }{\Delta^2+\gamma_A^2} + \dfrac{g_B^2 }{\Delta^2+\gamma_B^2}  \right). 
\end{align*}

\begin{figure}[btp]
\centering
\includegraphics[width=0.51\textwidth]{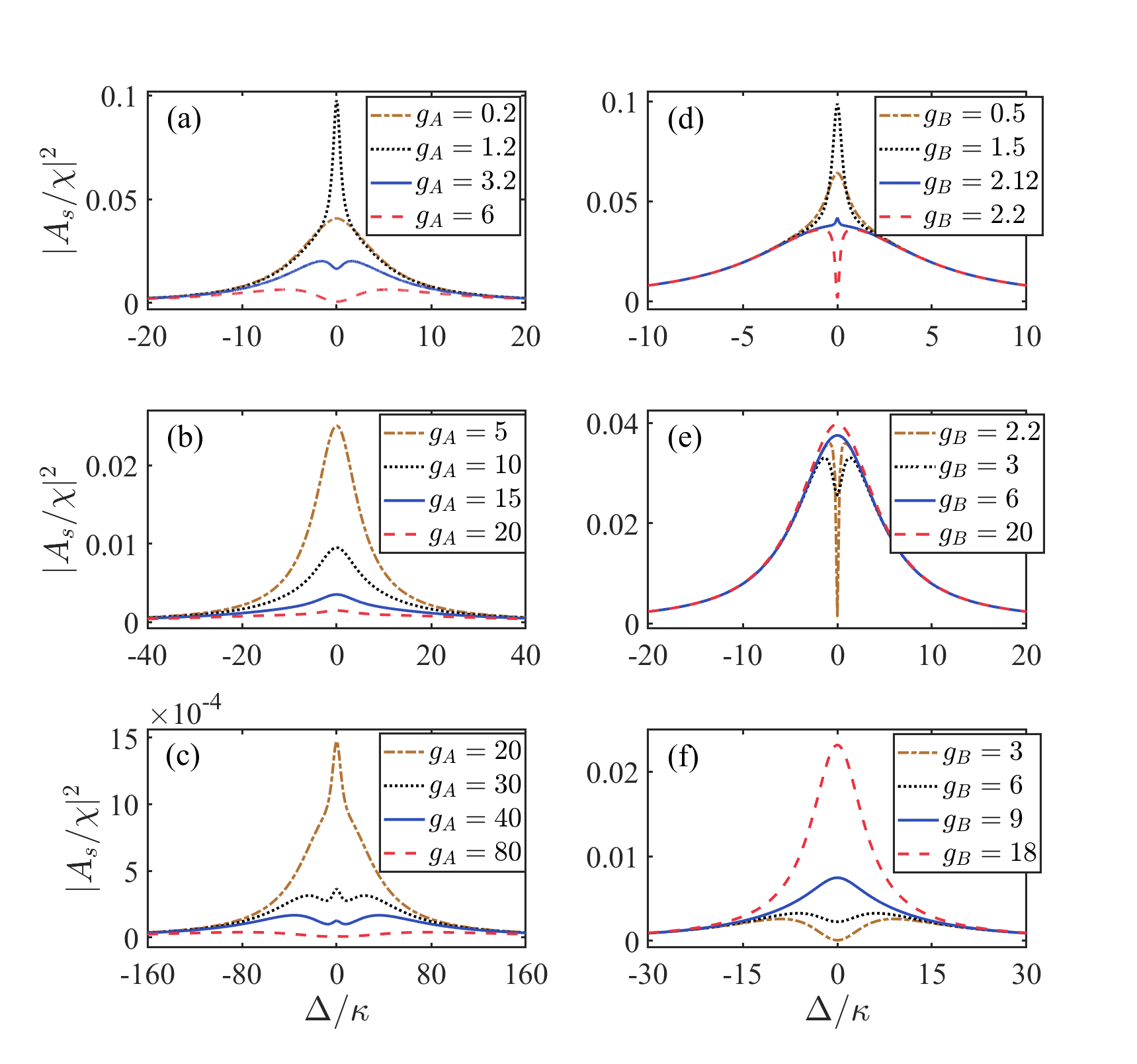} 
\caption{The response intensities of ion ensemble $\vert A_s/\chi \vert^2  $  as a function of the detuning  $ \Delta $ for (a) $ g_B=1 $, (b,c) $ g_B= 10 $ with different coupling strengths $ g_A $, or for (d,e) $ g_A= 1 $, (f) $ g_A= 10 $ with different coupling strengths $ g_B $. 
Here  $\gamma_A = \gamma_B = 5 $.  }
\label{VIA}
\end{figure}

In this case, the response spectra of the right ensemble always displays Lorentz-type peak without transparency window, while the left one shows some differences from the above case. In Fig. \ref{VIA}, we plot the response spectra of the driven ensemble as a function of the scaled detuning $ \Delta/\kappa $ for fixed decay rates. 
As shown in Fig. \ref{VIA}(a) with fixed $ g_B/\kappa= 1 $, the response intensity $ \vert A_s /\chi \vert^2 $ appears with a Lorentz peak, then VIA phenomenon emerges and the absorption peak increases with the raise of $ g_A $, but soon decreases and turns into one weak VIT window as $ g_A $ continuously increases.
From Fig. \ref{VIA}(b,c) with fixed $ g_B/\kappa= 10 $, it can be found that the Lorentz peak firstly evolves into a weak absorption peak with the increase of $ g_A $, then the central peak decreases and two weak transparency windows appear, which finally evolve into a weak VIT window with $ g_A $ becoming very large.  
In the condition of fixed $ g_A/\kappa= 1 $ in Fig. \ref{VIA}(d), the driven
ensemble responds with a VIA phenomenon where the absorption peak increases with the raise of $ g_B $.  Then the VIA peak quickly decreases and converts into single VIT window as $ g_B $ smoothly increases. With the continues increase of $ g_B $, the single VIT window eventually evolves into a Lorentz peak as shown in Fig. \ref{VIA}(e).  
From Fig. \ref{VIA}(f) with fixed $ g_A/\kappa= 10 $, one can find that 
the single VIT window becomes a Lorentz peak with the increase of $ g_B $.
Under proper conditions, the conversion from the absorption peak to the transparency window appears.

\section{Fluctuation spectrum}

To account for the effects of the quantum fluctuations we decompose the bosonic operators in the Langevin equations as the sum of their steady-state value and a small fluctuation, i.e., $ X =  X_s +\delta  X  $. 
The corresponding linear quantum Langevin equations for the fluctuations can be easily obtained.
For example, in the red-detuning case one  
can find 
\begin{align}
\delta\dot{c} =& -i \Delta \delta c-g_A\delta A-g_B\delta B-\kappa \delta c+\sqrt{2\kappa}c_{in}(t),  \\
\delta\dot{A} =& -i \Delta \delta A  + g_A \delta c-\gamma_A \delta A+ \sqrt{2\gamma_A}A_{in}(t),\\
\delta\dot{B} =& -i \Delta \delta B+g_B \delta c-\gamma_B \delta B + \sqrt{2\gamma_B}B_{in}(t).  
\end{align}

Without loss of generality, one can solve the Langevin equations conveniently in the frequency domain by taking the Fourier transform of each equation as
\begin{equation}
 \tilde{X}(\omega)= \dfrac{1}{2\pi} \int_{-\infty}^{+\infty}  X(t)e^{i\omega t}\,dt. \label{Fourier}
\end{equation}
It is easy to  find the solutions: 
\begin{align}
\delta\tilde{c}(\omega) =& \dfrac{1}{\omega-\Delta_{\text{eff}}(\omega)+i\kappa_{\text{eff}}(\omega)}\left[ i \sqrt{2\kappa}\tilde{c}_{in}(\omega)\right. 	\nonumber \\
+&\left.  \dfrac{g_A \sqrt{2\gamma_A}\tilde{A}_{in}(\omega)}{(\omega-\Delta)+i\gamma_A}	+ \dfrac{g_B \sqrt{2\gamma_B}\tilde{B}_{in}(\omega)}{(\omega-\Delta)+i\gamma_B} \right], 	\\
\delta\tilde{Y}(\omega) =& \dfrac{ig_Y \delta\tilde{c}(\omega)+i\sqrt{2\gamma_Y}\tilde{Y}_{in}(\omega)}{(\omega-\Delta)+i\gamma_Y} \quad (Y=A,B),	 
\end{align}
where
\begin{align*}
\Delta_{\text{eff}}(\omega) &=\Delta + \dfrac{g_A^2 (\omega-\Delta) }{ (\omega-\Delta)^2+\gamma_A^2} + \dfrac{g_B^2  (\omega-\Delta)}{ (\omega-\Delta)^2+\gamma_B^2},	\\
\kappa_{\text{eff}}(\omega) &=\kappa + \dfrac{g_A^2 \gamma_A}{ (\omega-\Delta)^2+\gamma_A^2} + \dfrac{g_B^2 \gamma_B}{ (\omega-\Delta)^2+\gamma_B^2} .
\end{align*}

\begin{figure}[bp]
\centering
\includegraphics[width=0.51\textwidth]{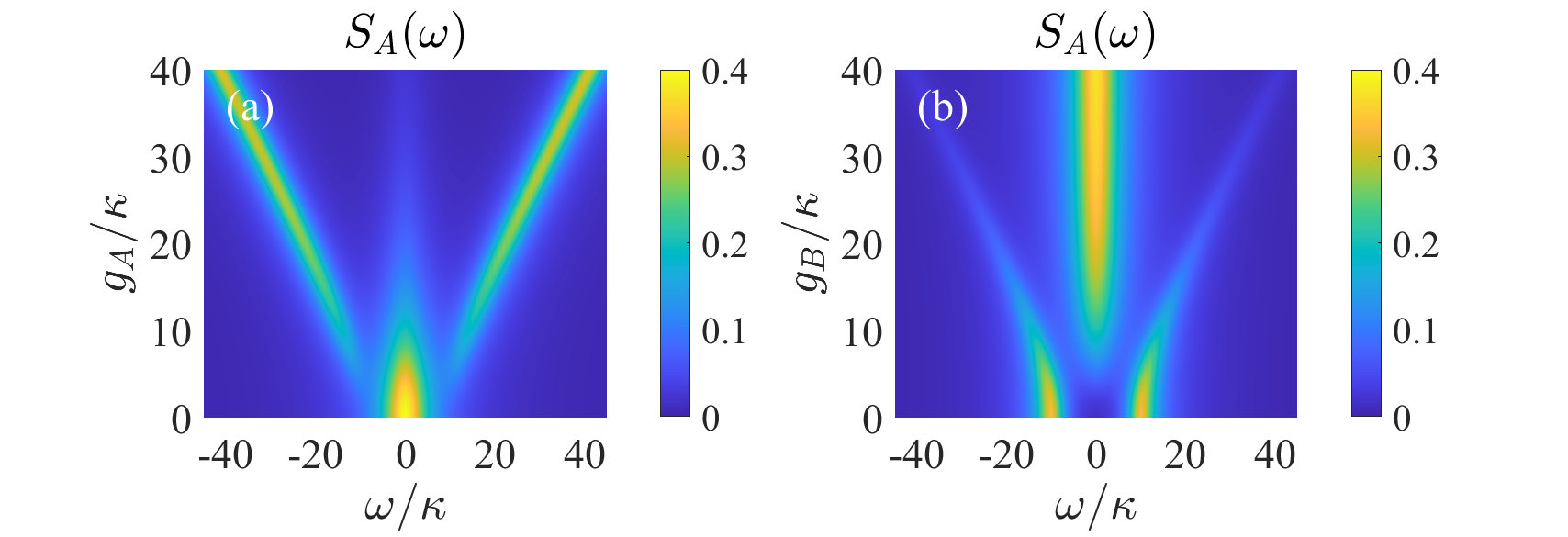} 
\caption{The fluctuation spectra of driven ion ensemble, $ S_A(\omega) $ in Eq. \eqref{SY}, as a function of  (a)  $ \omega/\kappa $  and   $ g_A/\kappa  $ for $ g_B = 10 $, (b) $ \omega/\kappa $  and $ g_B/\kappa $ for $ g_A = 10 $. Here $ \Delta=0 $ and $\gamma_A = \gamma_B = 5$ (in units of $ \kappa $). }
\label{fluct}
\end{figure}

Now we calculate the fluctuation spectra of the ionic vibrational mode and  collective-excitation modes, $ S_X(\omega) $ ($ X=c,A,B $), which are defined as \cite{Walls}
\begin{equation}
S_{X} (\omega)= \dfrac{1}{2\pi} \int_{-\infty}^{+\infty} \langle \delta X(t-\tau)  \delta X^{\dagger}(t) \rangle e^{i\omega \tau}\,d\tau .
\end{equation} 
The explicit forms of the fluctuation spectra of the collective-excitation modes for the ion ensemble are  
\begin{equation}
S_Y(\omega)= \dfrac{g_Y^2 S_c(\omega) + 2\gamma_Y [N(\omega_{eg})+1]\left[ 1+2g_Y^2 \Gamma_Y(\omega)\right] }{(\omega-\Delta)^2+\gamma_Y^2}, \label{SY}
\end{equation}
where
\begin{equation}
S_c(\omega)= \dfrac{2\kappa[N(\nu)+1]+2[N(\omega_{eg})+1] [ \kappa_{\text{eff}}(\omega)-\kappa  ] }{\left[ \omega-\Delta_{\text{eff}}(\omega)\right]^2 + \kappa^2_{\text{eff}}(\omega)}
\end{equation}
is the fluctuation spectrum of the ionic vibrational mode, and
\begin{equation*}
\Gamma_Y(\omega)=	\dfrac{(\omega-\Delta)\left[ \omega-\Delta_{\text{eff}}(\omega)\right]- \gamma_Y \kappa_{\text{eff}}(\omega)}{\left( \left[ \omega-\Delta_{\text{eff}}(\omega)\right]^2 + \kappa^2_{\text{eff}}(\omega) \right) \left[(\omega-\Delta)^2+\gamma_Y^2 \right]  }.
\end{equation*}

To investigate the properties of the fluctuation spectra of the left ion ensemble, Fig. \ref{fluct} displays the fluctuation spectra of $ S_A(\omega) $ as a function of the coupling strength  $ g_A $ or $ g_B $ for the fixed decay rates. 
For simplicity, here we consider the noise response spectrum with $ N(\nu)= N(\omega_{eg})=0$.
As shown in Fig. \ref{fluct}(a) where $ g_B=10\kappa $ and $\gamma_A = \gamma_B = 5\kappa$, we find the Lorentz central peak would evolve into a three-peak structure with two transparency windows as $ g_A $ increases, which is similar to the magnon-induced transparency and absorption effects \cite{BWang}.
With the continue increase of  $ g_A $, the central peak becomes narrower and lower, while the sidebands being two symmetrical peaks about $ \omega/\kappa= 0 $ deviate away from the central peak and become wide.
Whereas, as shown in Fig. \ref{fluct}(b) for fixed $ g_A=10\kappa $ (as well as $\gamma_A = \gamma_B = 5\kappa$), the fluctuation spectrum appears with one VIT window for weak $ g_B $, with two VIT windows for strong $ g_B $,  and with a Lorentz peak for ultra- or deep-strong $ g_B $. 
These agree with the previous analysis in Fig. \ref{VIT3}.

\begin{figure}[bp]
\centering
\includegraphics[width=0.51\textwidth]{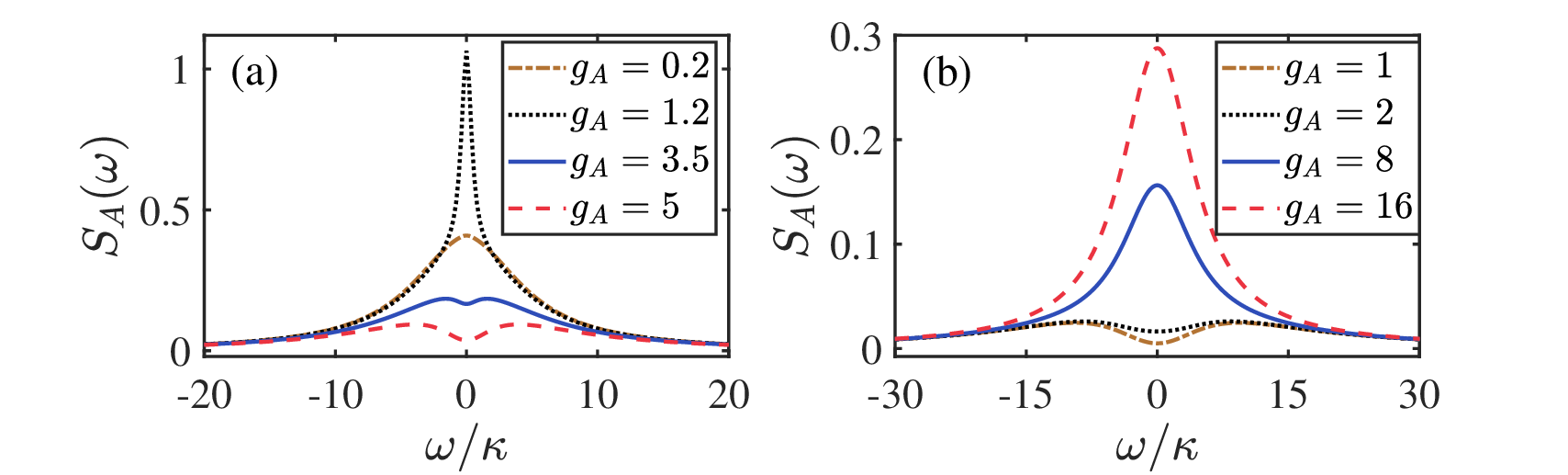} 
\caption{The fluctuation spectra of driven ion ensemble, $ S_A(\omega) $ in Eq. \eqref{SY'}, as a function of  $ \omega/\kappa $ for  (a) different coupling strengths   $ g_A $  with fixed $ g_B = 1 $ or (b) different coupling strengths $ g_B $  with fixed $ g_A = 10 $. For the other parameters, see Fig. \ref{fluct}. }
\label{fluct2}
\end{figure}

\bigskip  
In the blue-detuning case, following the previous steps one can obtain the explicit forms of the fluctuation spectra of the collective-excitation modes for the ion ensemble
\begin{align}
S'_Y(\omega)= \dfrac{g_Y^2 S'_{c^{\dagger}}(\omega) + 2\gamma_Y [N(\omega_{eg})+1]\left[ 1 -2g_Y^2 \Gamma'_Y(\omega)\right] }{(\omega-\Delta)^2+\gamma_Y^2}	,\label{SY'}
\end{align}
where
\begin{equation}
S'_{c^{\dagger}}(\omega)= \dfrac{2\kappa N(\nu) +2[N(\omega_{eg})+1] [\kappa- \kappa'_{\text{eff}}(\omega)  ] }{\left[ \omega-\Delta'_{\text{eff}}(\omega)\right]^2 + \kappa'^2_{\text{eff}}(\omega)} \label{Sc+'}
\end{equation}
and
\begin{align*}
\Delta'_{\text{eff}}(\omega) &=	2\Delta - \Delta_{\text{eff}}(\omega)	\\
&= \Delta - \dfrac{g_A^2 (\omega-\Delta) }{ (\omega-\Delta)^2+\gamma_A^2} - \dfrac{g_B^2  (\omega-\Delta)}{ (\omega-\Delta)^2+\gamma_B^2},	\nonumber\\
\kappa'_{\text{eff}}(\omega)&= 2\kappa -	\kappa_{\text{eff}}(\omega)	\\
&=\kappa - \dfrac{g_A^2 \gamma_A}{ (\omega-\Delta)^2+\gamma_A^2} - \dfrac{g_B^2 \gamma_B}{ (\omega-\Delta)^2+\gamma_B^2}, \nonumber\\
\Gamma'_{Y}(\omega) &= \dfrac{(\omega-\Delta)\left[ \omega-\Delta'_{\text{eff}}(\omega)\right]- \gamma_{Y} \kappa'_{\text{eff}}(\omega)}{\left( \left[ \omega-\Delta'_{\text{eff}}(\omega)\right]^2 + \kappa'^2_{\text{eff}}(\omega) \right) \left[(\omega-\Delta)^2+\gamma_{Y}^2 \right]  }.
\end{align*}

For simplicity,  we plot the fluctuation spectra $ S_A(\omega) $ as a function of $ \omega/\kappa $ for different coupling strength  $ g_A $ or $ g_B $ in Fig. \ref{fluct2}. 
As shown in Fig. \ref{fluct2}(a) where $ g_B=1\kappa $, we find the fluctuation spectra of the left ion ensemble, $ S_A(\omega) $, have the similar curves to the response intensity  $ \vert A_s /\chi \vert^2 $ in Fig. \ref{VIA}(a).
Likewise, in Fig. \ref{fluct2}(b) for fixed $ g_A=10\kappa $,  the fluctuation spectra $ S_A(\omega) $ are similar  to the curves  in Fig. \ref{VIA}(f).

\section{Discussion and conclusion}

\begin{figure}[btp]
\centering
\includegraphics[width=0.5\textwidth]{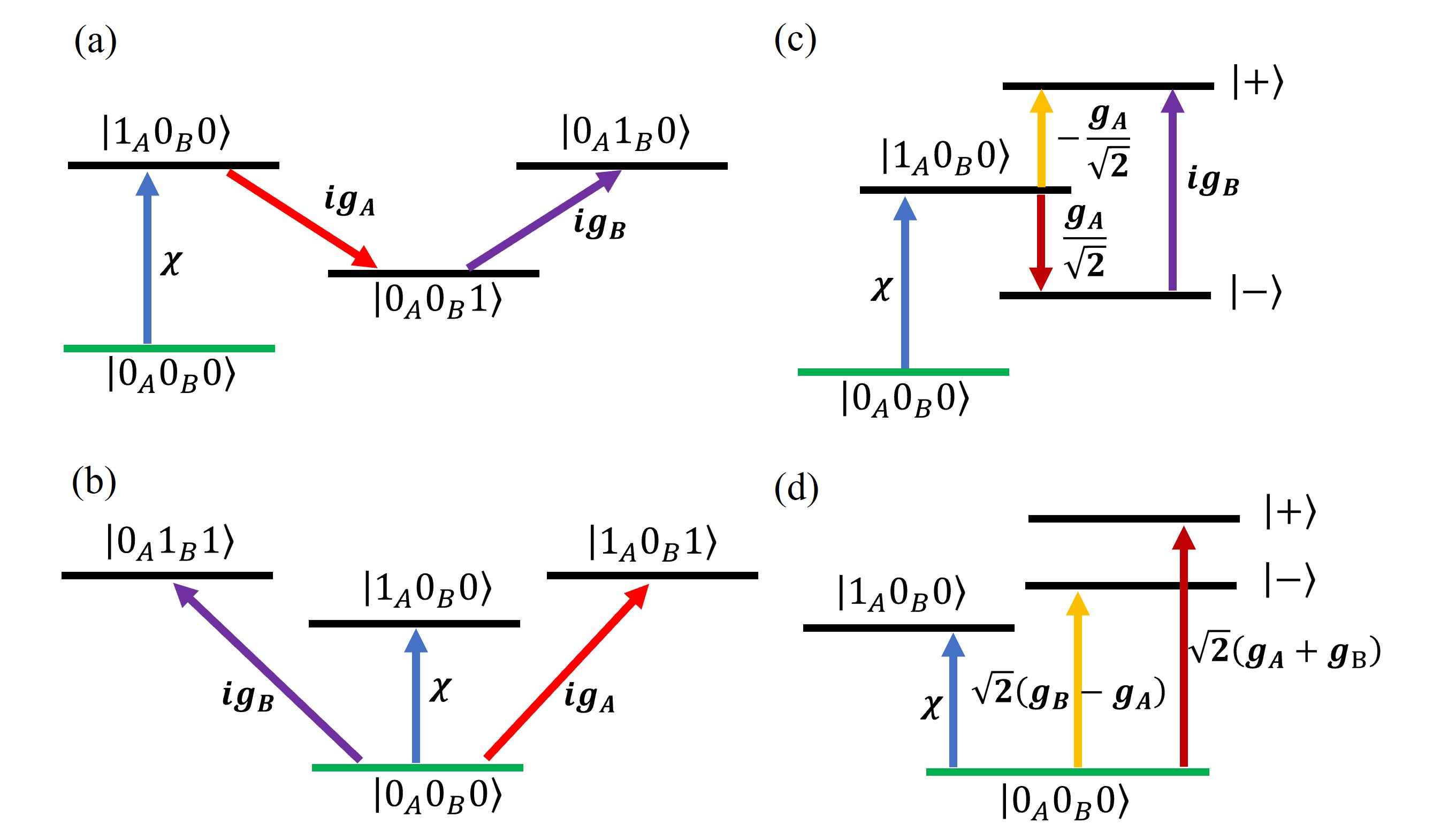} 
\caption{Energy level diagram of the model system in (a) red-detuning case and (b) blue-detuning case, with corresponding diagram (c) and (d) sketched by dressed states, in which $ \vert m_A \rangle$, $ \vert m_B \rangle$, and $ \vert n \rangle$ are number states of the two ionic collective-excitation modes and   vibrational mode. The coupling strengths are denoted by $ ig_A $ and $ ig_B $, respectively. The one-boson transition of the left collective mode is induced by the external probe field with driving strength $ \chi $. Also, in (c) dressed states $ \vert \pm \rangle = \sqrt{2}i(\vert 0_A 1_B 0 \rangle \pm  \vert 0_A 0_B 1 \rangle)/2$  and in (d) $ \vert \pm \rangle = \sqrt{2}i(\vert 0_A 1_B 1 \rangle \pm  \vert 1_A 0_B 1 \rangle)/2$.}
\label{path}
\end{figure}

In the consideration of the collective low excitations of two trapped ensembles and the small number states of ionic vibrational mode, the total states of the system in the red-detuning case are spanned by $\left\lbrace  \vert 0_A 0_B 0 \rangle, \vert 1_A 0_B 0 \rangle, \vert 0_A 1_B 0 \rangle ,  \vert 0_A 0_B 1 \rangle \right \rbrace $,  where $ \vert m_A  m_B  n \rangle = \vert m_A \rangle \otimes \vert m_B \rangle \otimes \vert n \rangle$. The energy level structure of these four states is sketched in Fig. \ref{path}(a). 
Similarly, in the blue-detuning case the total states of the system are spanned by  $\left\lbrace  \vert 0_A 0_B 0 \rangle, \vert 1_A 0_B 0 \rangle, \vert 0_A 1_B 1 \rangle ,  \vert 1_A 0_B 1 \rangle \right \rbrace $, and the energy level structure is sketched in Fig. \ref{path}(b). 
Replacing the states $ \vert 0_A 1_B 1 \rangle $  and $  \vert 1_A 0_B 1 \rangle $ with $ \vert \pm \rangle $, the transition processes sketched by dressed states is shown in Fig. \ref{path}(d).   
When $ g_A=0 $, it can be found that a Lorentz peak appears due to the destructive interference among three paths, with two of which (marked in yellow and red) have the same coupling strength. 
Whereas, for $ g_A \rightarrow g_B $, the second interference channel (marked in yellow) is suppressed, which is responsible for the transformation from VIA phenomena to VIT.
When $ g_A \gg g_B $, a weak VIT phenomenon is acquired as the coupling strengths of the two of three paths (marked in yellow and red) have approximate values but with opposite sign. 
However, in the red-detuning case the dressed states $ \vert \pm \rangle $, as shown in Fig. \ref{path}(c), replace the states $ \vert 0_A 1_B 0 \rangle $  and $  \vert 0_A 0_B 1 \rangle $. When $ g_A=0 $, four states form two separate paths (marked in blue and purple) without interference, which gives a Lorentz peak.   
When $ g_B =0 $, the VIT phenomena appears due to the quantum interference between  two paths (marked in red and yellow) with the coupling strength of same value but opposite sign, besides the isolated path (marked in blue).
And interference between three paths (marked in red, yellow and purple) gives rise to the generation of VIT phenomena for large value of parameters $ g_A$ and $ g_B $ or a Lorentz peak for small values of $ g_A $ and $ g_B $.

In conclusion, we investigate the generation of VIT and VIA for two atomic ion ensembles inside a liner trap when one of both is driven by a laser.
It is found that in the red-detuning case the VIT windows can be controlled by the amount coupling coefficients under certain conditions.
While in the blue-detuning case, we found that the VIT window can be converted into a VIA peak if we choose suitable values of the parameter $ g_A$ and  $ g_B $.
In addition, from Eqs. \eqref{SY} and \eqref{SY'}, it is easy to get that all the peaks of $ S_A(\omega) $ and $ S'_A(\omega)$  increase with the increase of temperature due to the noise being enhanced by higher environmental temperature \cite{FWang}.
We believe that our results would be useful for further understanding VIT and VIA and for exploring the potential applications in quantum information processing,  quantum circuit and chip, quantum communication, ect.

\section*{Acknowledgement} 
We thank Westlake University supercomputer center for the facility support and technical assistance.  
This work is supported by Natural Science Foundation of China (Grant No. 12574176), Wuxi Taihu University (2025THQD058) and Taihu Institute of Key Technologies for Integrated Circuits (TH25YB004).


\begin{thebibliography}{99}                                                                                               


\bibitem {Weiner} J. Weiner, P.-T. HO, 
\emph{LIGHT-MATTER INTERACTION}, Volume 1 Fundamentals and Applications (John Wiley \& Sons, Inc., Hoboken, New Jersey, 2003).

\bibitem {Kockum} A. F. Kockum, A. Miranowicz, S. D. Liberato, S. Savasta and F. Nori, Ultrastrong coupling between light and matter, 
\href{https://doi.org/10.1038/s42254-018-0006-2}{Nat. Rev. Phys. \textbf{1}, 19 (2019).}

\bibitem {Scully} M. O. Scully, M. S. Zubairy,
\emph{Quantum Optics} (Cambridge University Press, 1997).

 
\bibitem {Akulshin}  A. M. Akulshin, S. Barreiro, and A. Lezama, 
Electromagnetically induced absorption and transparency due to resonant two-field excitation of quasidegenerate levels in Rb vapor, 
\href{https://doi.org/10.1103/PhysRevA.57.2996}{Phys. Rev. A \textbf{57}, 2996 (1998).}


\bibitem {Lukin} M. D. Lukin \& A. Imamo\v{g}lu, 
Controlling photons using electromagnetically induced transparency,
\href{https://doi.org/10.1038/35095000}{Nature \textbf{413}, 273 (2001).}



\bibitem {EIT} M. M\"{u}cke, E. Figueroa, J. Bochmann, C. Hahn, K. Murr, S. Ritter, C. J. Villas-Boas \& G. Rempe, 
Electromagnetically induced transparency with single atoms in a cavity, 
\href{https://doi.org/10.1038/nature09093}{Nature \textbf{465},  755 (2010).}

\bibitem {EIT2} R. R\"{o}hlsberger, H.-C. Wille, K. Schlage \& B. Sahoo, 
Electromagnetically induced transparency with resonant nuclei in a cavity, 
\href{https://doi.org/10.1038/nature10741}{Nature 482, 199 (2012). }


\bibitem {EIA} A. Lezama, S. Barreiro, and A. M. Akulshin, 
Electromagnetically induced absorption, 
\href{https://doi.org/10.1103/PhysRevA.59.4732}{Phys. Rev. A \textbf{59}, 473 (1999).}

\bibitem {BWang} B. Wang, Z.-X. Liu, C. Kong, H. Xiong, and Y. Wu, 
Magnon-induced transparency and amplification in PT-symmetric cavity-magnon system,
\href{https://doi.org/10.1364/OE.26.020248}{Opt. Express \textbf{26}, 20248 (2018).}

\bibitem {Ullah} K. Ullah, M. T. Naseem, and \"{O}. E. M\"{u}stecapl\i\v{g}lu,
Tunable multiwindow magnomechanically induced transparency, Fano resonances, and slow-to-fast light conversion,
\href{https://doi.org/10.1103/PhysRevA.102.033721}{Phys. Rev. A \textbf{102}, 033721 (2020).}


\bibitem {FWang} F. Wang and C. Gou, 
Magnon-induced absorption via quantum interference,
\href{https://doi.org/10.1364/OL.482999}{Opt. Lett. \textbf{48}, 1164 (2023).}

\bibitem {Hou} B. P. Hou, L. F. Wei, and S. J. Wang, 
Optomechanically induced transparency and absorption in hybridized optomechanical systems, 
\href{https://doi.org/10.1103/PhysRevA.92.033829}{Phys. Rev. A \textbf{92}, 033829 (2015). }

\bibitem {Zhang} X. Y. Zhang, Y. H. Zhou, Y. Q. Guo, and X. X. Yi, 
Double optomechanically induced transparency and absorption in parity-time-symmetric optomechanical systems, 
\href{https://doi.org/10.1103/PhysRevA.98.033832}{Phys. Rev. A \textbf{98}, 033832 (2018).}

\bibitem {T. Wang} T. Wang, M.-H. Zheng, C.-H. Bai, D.-Y. Wang, A.-D. Zhu, H.-F. Wang, S. Zhang, 
Normal-Mode Splitting and Optomechanically Induced Absorption, Amplification, and Transparency in a Hybrid Optomechanical System, 
\href{https://doi.org/10.1002/andp.201800228}{Ann. Phys. \textbf{530}, 1800228 (2018).}
 
\bibitem {Aspelmeyer} M. Aspelmeyer, T. J. Kippenberg, and F. Marquardt, 
Cavity optomechanics, 
\href{https://doi.org/10.1103/RevModPhys.86.1391}{Rev. Mod. Phys. \textbf{86}, 1391 (2014).}


\bibitem {YU} S. YU, X. HE, P. XU, M. LIU, J. WANG, M. ZHAN,
Single atoms in the ring lattice for quantum information processing and quantum simulation,  
\href{https://doi.org/10.1007/s11434-012-5153-8}{Chinese Sci. Bull. \textbf{57}, 1931 (2012).}


\bibitem {Childs} A. M. Childs, J. Preskill, \& J. Renes,
Quantum information and precision measurement,
\href{https://doi.org/10.1080/09500340008244034}{J. Mod. Opt.  \textbf{47}, 155 (2000). }


\bibitem{Radeonychev} Y. V. Radeonychev, M. D. Tokman, A. G. Litvak, and O. Kocharovskaya,
Acoustically Induced Transparency in Optically Dense Resonance Medium,
\href{https://doi.org/10.1103/PhysRevLett.96.093602}{Phys. Rev. Lett. \textbf{96}, 093602 (2006).}

\bibitem {shaolpl} W. Shao, F. Wang, X.-L. Feng and C. H. Oh, 
Ionic vibration induced transparency and Autler-Townes splitting, 
\href{10.1088/1612-202X/aa621e}{Laser Phys. Lett. \textbf{14}, 045203 (2017).}


\bibitem {Turek} Y. Turek, Y. Li, and C. P. Sun, 
Electromagnetically-induced-transparency-like phenomenon with two atomic ensembles in a cavity, 
\href{https://doi.org/10.1103/PhysRevA.88.053827}{Phys. Rev. A \textbf{88}, 053827 (2013).}

\bibitem {Cirac} J. I. Cirac and P. Zoller, 
Quantum Computations with Cold Trapped Ions, 
\href{https://doi.org/10.1103/PhysRevLett.74.4091}{Phys. Rev. Lett. \textbf{74}, 4091 (1995).}
 

\bibitem {MS1} A. S\o rensen and K. M\o lmer, Quantum Computation with Ions in
Thermal Motion, 
\href{https://doi.org/10.1103/PhysRevLett.82.1971}{Phys. Rev. Lett. \textbf{82}, 1971 (1999).}

\bibitem {MS2} K. M\o lmer and A. S\o rensen, Multiparticle Entanglement of Hot
Trapped Ions, 
\href{https://doi.org/10.1103/PhysRevLett.82.1835}{Phys. Rev. Lett. \textbf{82}, 1835 (1999).}

\bibitem {Leibfried} D. Leibfried, R. Blatt, C. Monroe, and D. Wineland, Quantum dynamics of single trapped ions, 
\href{https://doi.org/10.1103/RevModPhys.75.281}{Rev. Mod. Phys. \textbf{75}, 281 (2003).}

\bibitem {Liu} N. Liu, R. Chang, S. Zhang, J.-Q. Liang, 
Ground-State Cooling of the Mechanical Resonator in an Optomechanical Cavity with Two-Level Atomic Ensemble, 
\href{https://doi.org/10.1007/s10773-022-05103-z}{Int. J. Theor. Phys. \textbf{61}, 120 (2022).}

\bibitem {Jin} G. R. Jin, P. Zhang, Y.-x. Liu, and C. P. Sun, 
Superradiance of low-density Frenkel excitons in a crystal slab of three-level atoms: The quantum interference effect, 
\href{https://doi.org/10.1103/PhysRevB.68.134301}{Phys. Rev. B \textbf{68}, 134301 (2003).}

\bibitem {Gardiner} C. Gardiner and P. Zoller, \emph{Quantum Noise} (Springer Science \& Business Media, 2004).
 
\bibitem {Walls} D. F. Walls and G. J. Milburn, \emph{Quantum Optics} (Springer, Berlin, 1994).



\end{thebibliography}
\end{document}